\documentclass[a4paper,11pt]{article}
\pdfoutput=1 

\usepackage{jheppub} 

\usepackage{multirow}
\usepackage{nicefrac}
\newcommand{\ba}{\begin{eqnarray}}
\newcommand{\ea}{\end{eqnarray}}

\usepackage[T1]{fontenc} 

\title{\boldmath Further refining Swampland Conjecture on inflation in general scalar-tensor theories of gravity}

\author[a]{Jureeporn Yuennan,}
\author[b,c]{Phongpichit Channuie}

\affiliation[a]{Surface Technology Research Unit (STRU), Faculty of Science and Technology, \\Nakhon Si Thammarat Rajabhat University, Nakhon Si Thammarat, 80280, Thailand}
\affiliation[b]{College of Graduate Studies, Walailak University, Thasala, Nakhon Si Thammarat, \\80160, Thailand}
\affiliation[c]{School of Science, Walailak University, Thasala, Nakhon Si Thammarat, \\80160, Thailand}

\emailAdd{jureeporn\_yue@nstru.ac.th, channuie@gmail.com}

\abstract{An alternative refined de Sitter conjecture giving rise to a natural combination of the first and second derivatives of the scalar potential was proposed recently by David Andriot and Christoph Roupec (Fortsch. Phys. {\bf 67} (2019) no.1-2, 1800105). In this work, we study the inflation models in a general scalar-tensor theory with exponential and hyperbolic tangent forms of potential as well as model with quantum corrected potential and examine whether these three models of inflation can satisfy this further refining de Sitter swampland conjecture or not. Regarding our analysis with proper choices of parameters  $a,\,b=1- a$ and $q$, we find that these three inflationary models can always satisfy this new refined swampland conjecture. Therefore, all three inflationary models might all be in “landscape” since the “further refining de Sitter swampland conjecture” is satisfied.}


\begin{document} 
\maketitle
\flushbottom

\section{Introduction}
In string theory, the swampland program has been recently developed for the phenomenology of quantum gravity theories. It connects to questions regarding effective field theories (EFTs) and their UV completions. Terminologically, the swampland is defined as the space of effective theories that cannot be consistently coupled to a theory of quantum gravity, and hence it includes the set of phenomenological models which cannot be derived as a low energy effective theory of a quantum gravity in the high energy regime. Therefore, it is required for consistent
EFTs not to lie in the swamplands, see a comprehensive review on the Swampland \cite{Brennan:2017rbf,Palti:2019pca},

Recently, the refined version of the swampland conjecture has been suggested \cite{Garg:2018reu,Ooguri:2018wrx}. In various phenomenological models \cite{Wang:2018kly,Fukuda:2018haz} including inflation \cite{Kinney:2018nny,Lin:2018rnx,Cheong:2018udx} and dark energy \cite{Agrawal:2018rcg,Chiang:2018lqx,Colgain:2019joh,Banerjee:2020xcn}, this refined
de Sitter conjecture has been tested. Moreover it has
been also discussed in relation to stringy constructions \cite{Olguin-Trejo:2018zun,Garg:2018zdg,Blaback:2018hdo,Heckman:2018mxl,Blanco-Pillado:2018xyn,Junghans:2018gdb,Emelin:2018igk,Banlaki:2018ayh} or in a more general swampland context \cite{Hebecker:2018vxz,Dvali:2018jhn,Schimmrigk:2018gch,Ibe:2018ffn}. More concretely, following a setup given in Ref.\cite{Andriot:2018mav}, we consider a four-dimensional ($4D$) theory of real scalar field $\phi^{i}$ coupled to gravity, and its dynamics is governed by a scalar potential $V(\phi^{j})$. Hence the
action can be written as
\begin{eqnarray}
S = \int_{4D} d^4x\sqrt{-g}\,\bigg[-\frac{1}{2}\,M_p^2\, R + \frac{1}{2}\,g^{\mu\nu}h_{ij}\partial_\mu \phi^{i}\,\partial_\nu \phi^{j} - V\bigg],\label{Ac}
\end{eqnarray}
where $h_{ij}(\phi^k)$ is the field space metric, $M_{p}$ is the 4d Planck mass, and the $4D$ space-time indexes $(\mu,\,\nu)$ are raised and lowered with the $4D$ metric $g_{\mu\nu}$ with a signature $(+,-,-,-)$. Various phenomenological models, such as multi-field cosmological inflation \cite{Dimopoulos:2005ac,Wands:2007bd}, can be described by the above action. Regarding the refined de Sitter conjecture, an effective theory for quantum gravity, i.e. not in the swampland, should satisfy one of the following two conditions \cite{Garg:2018reu,Ooguri:2018wrx}:
\begin{eqnarray}
|\nabla V|\geq \frac{c_{1}}{M_{p}}V\,,\label{R1}
\end{eqnarray}
or
\begin{eqnarray}
{\rm min}(\nabla_{j}\nabla_{j} V)\,\,\leq -\frac{c_{2}}{M^{2}_{p}}V\,,\label{R2}
\end{eqnarray}
where $c_1$ and $c_2$ are both positive constants of the order of ${\cal O}(1)$ and $|\nabla V|=\sqrt{g^{ij}\nabla_{j}\nabla_{j} V}$. Therefore,  for any $V$, the standard slow-roll parameters can be recast using the inequalities to yield
\begin{eqnarray}
\sqrt{2\epsilon_{V}}\geq c_{1}\,,\quad{\rm or}\quad \eta_{V}\,\,\leq -c_{2}\,.\label{R12}
\end{eqnarray}
The first condition corresponds to the original
“swampland conjecture” proposed in Ref.\cite{Obied:2018sgi}. However, a peculiarity of this conjecture regarding these two distinct conditions (\ref{R1}) and (\ref{R2}) on two different quantities $\epsilon_{V}$ and $\eta_{V}$ was noticed. Based on this discussion, a single condition on both $\epsilon_{V}$ and $\eta_{V}$ has been proposed. The authors named it as a further
refining de Sitter swampland conjecture \cite{Andriot:2018mav}. 

The statement of an alternative refined de Sitter conjecture is suggested that a low energy effective theory of a quantum gravity that takes the form (\ref{Ac}) should verify, at any point in field space where $V>0$ \cite{Andriot:2018mav},
\begin{eqnarray}
\Big(M_{p}\frac{|\nabla V|}{V}\Big)^{q}-aM^{2}_{P}\frac{{\rm min}(\nabla_{j}\nabla_{j} V)}{V}\geq b\,\quad{\rm with}\quad a+b=1,\,a,\,b>0,\,q>2\,,\label{ReRe}
\end{eqnarray}
which gives a natural condition on a combination of
the first and second derivatives of the scalar potential. In terms of the slow-roll parameters, the conjecture can be rewritten as [1]:
\begin{eqnarray}
(2\epsilon_{V})^{q/2}-a\eta_{V}\geq b\,.\label{Re1}
\end{eqnarray}
Interestingly, the authors of Ref.\cite{Liu:2021diz} have examined if Higgs inflation model, Palatini Higgs inflation, and Higgs-Dilaton model can satisfy the further refining de Sitter swampland conjecture or not, and it is found that these three inflationary models can always satisfy a new swampland conjecture if only they adjust the relevant parameters $a,\,b=1-a$ and $q$. 

In the present work, this interesting and concrete conjecture will be tested in  inflation in general scalar-tensor theories of gravity. The paper is organized in the following way. In Section \ref{s2}, we briefly review the inflation models in a general scalar-tensor theory including inflation with exponential and tangential forms of potential as well as model with quantum corrected potential. For each model, the spectral index and the tensor-to-scalar ration will be derived. In Section \ref{s3}, we examine whether these
models satisfy the further refining swampland conjecture or not. Finally, we conclude our findings in the last section. 

\section{Inflationary models in general scalar-tensor theory}
\label{s2}
In this section, we briefly review the inflation models in a general scalar-tensor theory: inflation with E-form potential, T-form potential, and  the quantum  corrected potential. We start our study with the action of a general scalar-tensor theory in the Jordan frame and it takes the form \cite{Amake:2021bee}
\begin{eqnarray}
S^{J} = \int d^4x\sqrt{-g}\,\bigg[ -\frac{1}{2}\Omega^{2}(\phi)\,M_p^2\, R + \frac{1}{2}\omega(\phi)\,g^{\mu\nu}\partial_\mu \phi\,\partial_\nu \phi - V^{J}(\phi)\bigg],
\label{LJ}
\end{eqnarray}
where a superscript $J$ stands for quantities in the Jordan frame and the reduced Planck mass is defined as $M_p^2 = 1/8\pi G$. Here $\Omega^{2}(\phi)$ is given by
\begin{eqnarray}
\Omega^{2}(\phi) = \frac{M^{2}_{p}+\xi f(\phi)}{M^{2}_{p}},
\label{Ome}
\end{eqnarray}
where $f(\phi)$ is an arbitrary function on the scalar field $\phi$ and $\xi$ is a dimensionless coupling constant. By applying the conformal transformation ${\bar g}_{\mu\nu} = \Omega^{2}(\phi)g_{\mu\nu}$, we can eliminate the non-minimal coupling between $f(\phi)$ and the gravitational field. The resulting action in the Einstein frame reads
\begin{eqnarray}
S^{E} = \int d^4x\sqrt{-g}\,\bigg[ -\frac{1}{2}\,M_p^2\, R + \frac{1}{2}\,g^{\mu\nu}\partial_\mu \psi\,\partial_\nu \psi - U^{E}(\psi)\bigg]\,,
\label{LE}
\end{eqnarray}
where $U^{E}(\psi)=V^{J}(\phi)/\Omega^{4}(\phi)$, and a superscript $E$ stands for quantities in the Einstein frame and
\begin{eqnarray}
d\psi^{2} = \bigg[\frac{\omega(\phi)}{\Omega^{2}(\phi)} + 6M^{2}_{p}\frac{\Omega^{'2}(\phi)}{\Omega^{2}(\phi)} \bigg]\,d\phi^{2},\,
\label{Omgg}
\end{eqnarray}
If a conformal factor $\Omega(\phi)$ and a kinetic coupling $\omega(\phi)$ satisfy the condition
\begin{eqnarray}
\omega(\phi)=\frac{M^{2}_{p}}{\xi}\Omega^{'2}(\phi)\,,
\label{cons}
\end{eqnarray}
then there exists an exact relationship between $\phi$ and $\psi$ obtained from Eq.(\ref{Omgg}):
\begin{eqnarray}
\psi=\sqrt{6\alpha}\,M_{p}\ln \Omega,\,\,\,\,\,\,\Omega(\phi)=e^{\sqrt{1/6\alpha}\,\psi/M_{p}}\,.
\label{conso}
\end{eqnarray}
and
\begin{eqnarray}
V^{J}(\phi)=\Omega^{4}(\phi)U^{E}\big(\sqrt{6\alpha}\,M_{p}\ln \Omega(\phi)\big)\,,
\label{conso1}
\end{eqnarray}
where $\alpha=1+(6\xi)^{-1}$. In order to obtain the action in the Einstein frame, the following identities in 4 spacetime dimensions \cite{Fujii2003} are necessary:
\begin{eqnarray}
{\bar R} &=& \frac{1}{\Omega^{2}}\bigg[R-6\,g^{\mu\nu}\nabla_{\mu}\nabla_{\nu}\ln\phi - 6\,g^{\mu\nu}(\partial_\mu \ln\phi)(\partial_\nu \ln\phi)\bigg]\,,\nonumber\\
{\bar g}^{\mu\nu} &=& \Omega^{-2}g^{\mu\nu},\,\,\,\sqrt{-{\bar g}}=\Omega^{4}\sqrt{-g}\nonumber\,,
\label{ids}
\end{eqnarray}
where an argument of $\Omega$ is understood, and a bar denotes quantities in the Einstein frame, and we have omitted bars from here for convenience. 

\subsection{Inflation with exponential form of potential}
In the first case scenario, we consider the condition (\ref{Omgg}). If we take $V^{J}(\phi)=V_{0}\big(1-\Omega^{2}(\phi)\big)^{2}$, then we obtain the exponential form of the potential, named E-model, given in the Einstein frame of the form
\begin{eqnarray}
U^{E}(\psi)=V_{0}\bigg(1-e^{-2/\sqrt{6\alpha}\,\psi/M_{p}}\bigg)^{2}\,.
\label{conso11}
\end{eqnarray}
The associated conventional slow-roll parameters are given by
\begin{eqnarray}
\varepsilon = \frac{M_p^2}{2}\left( \frac{U_\psi}{U}\right)^2\,, \quad\quad\eta = M_p^2\,\frac{U_{\psi\psi}}{U}\,,
\label{SR-parameters}
\end{eqnarray}
where $U_{\psi}$ denotes derivative with respect to $\psi$, i.e., $U_\psi=dU/d\psi$ and $U_{\psi\psi}=d^{2}U/d\psi^{2}$. We obtain
\begin{eqnarray}
\varepsilon =\frac{4}{3 \alpha  \left(e^{\frac{\sqrt{\frac{2}{3}} \psi }{\sqrt{\alpha }M_{p}}}-1\right)^2}\,,\quad\quad \eta=-\frac{4 \left(e^{\frac{\sqrt{\frac{2}{3}} \psi }{\sqrt{\alpha } M_{p}}}-2\right)}{3 \alpha  \left(e^{\frac{\sqrt{\frac{2}{3}} \psi }{\sqrt{\alpha } M_{p}}}-1\right)^2}\,.\label{eeta}
\end{eqnarray}
Inflation ends when $\varepsilon=1$ and we find
\begin{eqnarray}
\psi_{\rm end}= \sqrt{\frac{3\alpha}{2}}M_{p}\log \left(\frac{2}{\sqrt{3\alpha}}+1\right)\,.\label{mo1}
\end{eqnarray}
The number of {\it e}-foldings during inflation is defined via
\begin{eqnarray}
N =\frac{1}{M^{2}_{p}}\int^{\psi_{\rm ini}}_{\psi_{\rm end}} \frac{U}{U_{\psi}}d\psi\simeq \frac{3 \alpha}{4}\left(e^{\frac{\sqrt{\frac{2}{3}} \psi_{\rm{ini}}}{\sqrt{\alpha } M_{p}}}-e^{\frac{\sqrt{\frac{2}{3}} \psi _{\rm{end}}}{\sqrt{\alpha} M_{p}}}\right)\,.
\end{eqnarray}
The above result can be combined with Eq.(\ref{mo1}) allowing us to write:
\begin{eqnarray}
\psi_{\rm ini}\simeq \sqrt{\frac{3\alpha}{2}}M_{p}\log \left(\frac{4 N}{3 \alpha }\right)\,.\label{ini}
\end{eqnarray}
To generate the proper amplitude of the density perturbations the potential must satisfy at WMAP the normalization
condition \cite{Planck:2018jri}:
\begin{eqnarray}
\frac{U}{\varepsilon}\Big|_{\psi_{\rm ini}}=(0.0276 M_{p})^{4}\quad\rightarrow\quad \frac{4 N^2 V_{0} \left(1-\frac{3\alpha}{4 N}\right)^4}{3\alpha}=(0.0276 M_{p})^{4}
\end{eqnarray}
corresponding to the initial value assumed by the inflaton. We therefore obtain
\begin{eqnarray}
V_{0}\simeq \frac{4.35\times 10^{-7} \alpha M_p^4}{N^2 \left(1-\frac{0.75 \alpha }{N}\right)^4}\,.
\end{eqnarray}
Ii is useful to write $\varepsilon$ and $\eta$ in terms of the number of e-foldings. Substituting Eq.(\ref{ini}) into Eq.(\ref{eeta}), we have
\begin{eqnarray}
\varepsilon = \frac{3 \alpha }{4 N^2}+\frac{9 \alpha ^2}{8 N^3}+O\left(\alpha ^3\right)\,,\quad
\eta = -\frac{1}{N}+\frac{9 \alpha ^2}{16 N^3}+O\left(\alpha ^3\right)\,.
\end{eqnarray}
The spectral index of curvature perturbation $n_s$ and the tensor-to-scalar
ratio $r$ are given in terms of the e-foldings $N$:
\begin{eqnarray}
n_{s} &=& 1-6\varepsilon+2\eta = \left(1-\frac{2}{N}\right)-\frac{9 \alpha }{2 N^2}+O\left(\alpha ^2\right)\label{ns1}\,,\\ r  &=& 16\varepsilon =\frac{12 \alpha }{N^2}+\frac{18 \alpha ^2}{N^3}+O\left(\alpha ^3\right)\,.\label{r1}
\end{eqnarray}
Note that the above results still hold since $N\gg\alpha$. Setting $\alpha=1$, we obtain the same results to those of the models found in the existing references including Higgs inflation \cite{Bezrukov:2007ep}, Starobinsky inflation \cite{Starobinsky:1980te}, Higgs Starobinsky inflation \cite{Calmet:2016fsr,Salvio:2015kka,Salvio:2016vxi}, and even composite models of inflation \cite{Channuie:2015ewa,Channuie:2016iyy}. 

\subsection{Inflation with hyperbolic tangent form of potential}
In the second model, under the condition (\ref{Omgg}), if we choose
\begin{eqnarray}
V^{J}(\phi)=V_{0}\Omega^{4}(\phi)\bigg(\frac{1-\Omega^{2}(\phi)}{1+\Omega^{2}(\phi)}\bigg)^{2}\,,
\label{cot}
\end{eqnarray}
then we get the hyperbolic tangent form of the potential, named it as T-model, in the Einstein frame
\begin{eqnarray}
U^{E}(\psi)=V_{0}\tanh^{2}\bigg(\frac{\psi}{\sqrt{6\alpha}M_{p}}\bigg)\,.
\label{cot2}
\end{eqnarray}
The associated slow-roll parameters are given by
\begin{eqnarray}
\varepsilon = \frac{M_P^2}{2}\left( \frac{U^{E}_\psi}{U^{E}}\right)^2\,, \quad\quad\eta = M_P^2\,\frac{U^{E}_{\psi\psi}}{U^{E}}\,,
\label{SR-parametersT}
\end{eqnarray}
where $U^{E}_{\psi}$ denotes derivative with respect to $\psi$, i.e., $U^{E}_\psi=dU^{E}/d\psi$ and $U^{E}_{\psi\psi}=d^{2}U^{E}/d\psi^{2}$. We obtain
\begin{eqnarray}
\varepsilon =\frac{4 \text{csch}^2\left(\frac{\sqrt{\frac{2}{3}}\psi}{\sqrt{\alpha}M_{P}}\right)}{3 \alpha }\,,\quad\quad \eta=\frac{\left(\text{csch}^2\left(\frac{\psi }{\sqrt{6} \sqrt{\alpha } M_{p}}\right)-2\right) \text{sech}^2\left(\frac{\psi }{\sqrt{6} \sqrt{\alpha } M_{p}}\right)}{3 \alpha }\,.\label{eetaT}
\end{eqnarray}
Inflation ends when $\varepsilon=1$ and we find
\begin{eqnarray}
\psi_{\rm end}= \sqrt{\frac{3\alpha}{2}} M_{p} \sinh^{-1}\left(\frac{2}{\sqrt{3\alpha}}\right)\,.\label{mo1T}
\end{eqnarray}
The number of {\it e}-foldings during inflation is defined via
\begin{eqnarray}
N =\frac{1}{M^{2}_{p}}\int^{\psi_{\rm ini}}_{\psi_{\rm end}} \frac{U}{U_{\psi}}d\psi\simeq \frac{3}{4} \alpha  \Big(\cosh \Big(\sqrt{\frac{2}{3\alpha}}\frac{\psi_{\text{ini}}}{M_{p}}\Big)-\cosh \Big(\sqrt{\frac{2}{3\alpha}}\frac{\psi_{\text{end}}}{M_{P}}\Big)\Big)\,.
\end{eqnarray}
The above result can be combined with Eq.(\ref{mo1T}) allowing us to write:
\begin{eqnarray}
\psi_{\rm ini}\simeq \sqrt{\frac{3\alpha}{2}}M_{p}\cosh ^{-1}\left(\frac{4 N}{3 \alpha }\right)\,.\label{iniT}
\end{eqnarray}
To generate the proper amplitude of the density perturbations the potential must satisfy at WMAP the normalization
condition \cite{Planck:2018jri}:
\begin{eqnarray}
\frac{U}{\varepsilon}\Big|_{\psi_{\rm ini}}=(0.0276 M_{P})^{4}\quad\rightarrow\quad \frac{V_{0} (4N-3\alpha )^2}{12 \alpha }=(0.0276 M_{P})^{4}
\end{eqnarray}
corresponding to the initial value assumed by the inflaton. We therefore obtain
\begin{eqnarray}
V_{0}\simeq \frac{6.96\times 10^{-6}\alpha M_{p}^4}{(4N-3\alpha )^2}\,.
\end{eqnarray}
Ii is useful to write $\varepsilon$ and $\eta$ in terms of the number of e-foldings. Substituting Eq.(\ref{ini}) into Eq.(\ref{eeta}), we have
\begin{eqnarray}
\varepsilon = \frac{3\alpha}{4 N^2}+O\left(\alpha^3\right)\,,\quad
\eta = -\frac{1}{N}+\frac{3\alpha}{2 N^2}-\frac{9\alpha^2}{16 N^3}+O\left(\alpha^3\right)\,.
\end{eqnarray}
The spectral index of curvature perturbation $n_s$ and the tensor-to-scalar
ratio $r$ are given in terms of the e-foldings $N$:
\begin{eqnarray}
n_{s} &=& 1-6\varepsilon+2\eta = \left(1-\frac{2}{N}\right)-\frac{3 \alpha }{2 N^2}-\frac{9 \alpha ^2}{8 N^3}+O\left(\alpha ^3\right)\,,\\ r  &=& 16\varepsilon =\frac{12 \alpha }{N^2}+O\left(\alpha ^3\right)\,.
\end{eqnarray}
With the results given above, there was a class of inflationary models so called cosmological $\alpha$-attractors has recently received considerable attention \cite{Kallosh:2013yoa,Kallosh:2014rga,Kallosh:2015lwa,Roest:2015qya,Linde:2016uec,Terada:2016nqg,Ueno:2016dim,Odintsov:2016vzz}. Note that the above results still hold since $N\gg\alpha$.

\subsection{Quantum corrected inflation}
We start with a action of the scalar field non-minimally coupled to gravity with a general form of the effective potential $V(\phi)$. It takes the form
\begin{eqnarray}
S^J = \int \sqrt{-g}\left[ -\frac12\left(M_p^2 + \xi \,\phi^2 \right)R + g^{\mu\nu}\partial_\mu\phi\partial_\nu\phi - V(\phi)\right]
\label{action-J}
\end{eqnarray}
where the action $S_J$ stands for the gravitational action in the Jordan frame. While $M_P^2\equiv 1/8\pi G$ and $\xi$ are reduced Plank mass and the non-minimal coupling constant, respectively. It is more convenient to study the inflation dynamics of the non-minimal coupling in the Einstein frame, i.e., the gravitational sector of the action written in the Einstein-Hilbert form only. The Einstein frame can be achieved by using the conformal transformation via a re-defining metric tensor as,
\begin{eqnarray}
{\bar g}_{\mu\nu} = \Omega(\phi)^2\,g_{\mu\nu}\,, \qquad \Omega(\phi)^2 = \frac{M_p^2 + \xi\,\phi^2}{M_p^2}\,.
\end{eqnarray}
Here all variables with tilde symbol represent the quantities in the Einstein frame and $f(\phi)=\phi^{2}$. Applying the conformal transformation to the action (\ref{action-J}), the action in Einstein frame is given by,
\begin{eqnarray}
S^E = \int \sqrt{-{\bar g}}\left[ -\frac12\,M_p^2\,{\bar g}^{\mu\nu}{\bar R}_{\mu\nu} + {\bar g}^{\mu\nu}\partial_\mu\psi\partial_\nu\psi - U(\psi)\right].
\label{action-E}
\end{eqnarray}
We have used the re-definition of new scalar field, $\psi$ in the Einstein frame to obtain the canonical form of the kinetic term of the scalar field, $\psi$ as
\begin{eqnarray}
\frac12\left( \frac{d\psi}{d\phi}\right)^2 = \frac{1+ 3\,M_p^2\,\Omega_\phi^2}{\Omega^2} \,,\label{mod2}
\end{eqnarray}
where $\Omega_\phi\equiv d \Omega/d\phi$. Comparing Eq.(\ref{mod2}) with Eq.(\ref{Omgg}), we find $\omega(\phi)=2$. The new effective potential in the Einstein frame, $U(\psi)$ is also given by,
\begin{eqnarray}
U(\psi) = \Omega^{-4}\,V\left(\phi(\psi)\right).
\label{E-potential}
\end{eqnarray}
In this section, we will consider the self-interacting potential with phenomenological quantum correction and this potential has been proposed by authors of Ref.\cite{Joergensen:2014rya} in order to analyze the characters of the quantum correction in the self-interacting scalar field phenomenology. They wrote the potential in the Jordan frame given by
\begin{eqnarray}
V(\phi) = \lambda\,\phi^4\left( \frac{\phi}{\Lambda}\right)^{4\gamma}\,.
\label{quantum-potential}
\end{eqnarray}
Here the quantum correction (real) parameter $\gamma$ is introduced and will be used to characterize the quantum behavior of the self-interacting potential. The $\Lambda$ parameter is the cut-off at a given energy scale. It was shown that the range of the $\gamma$ should be $\mathcal{O}(\gamma) \sim 0.1$ according to the constraint from observational data \cite{Joergensen:2014rya}. 

Before calculating the slow-roll parameters, we would like to express the form of the effective potential in the Einstein in Eq.(\ref{E-potential}) under the large field assumption during the inflation i.e., $ \phi \gg {M_P /\sqrt{\xi}}$\,. One finds,
\begin{eqnarray}
\psi \simeq \kappa\, M_p \ln \Big({ \sqrt{\xi}\, \phi \over  M_p } \Big) , ~~~~ \kappa \equiv \sqrt{{2\over \xi} + 6} \label{psh}
\end{eqnarray}
Then the Einstein frame potential then takes the following form 
\begin{eqnarray}
U(\psi) = \Omega^{-4} V(\phi(\psi)) &=& \frac{ M_p^4}{\left(M_p^2+\xi  \phi ^2\right)^2} \lambda\,  \phi ^4  \left(\frac{\phi }{\Lambda }\right)^{4 \gamma }
\nonumber\\
&=& {\lambda M_p^4 \over \xi^2} \left(\exp\Bigg[{\frac{-2 \psi}{\kappa M_p}} \Bigg] +1 \right)^{-2}  \left({M_p \over \sqrt{\xi} \Lambda} \right)^{4 \gamma} \exp \Bigg[{4 \gamma \psi \over \kappa M_p} \Bigg]\,. 
\label{U-chi}
\end{eqnarray} 
The associated slow-roll parameters are given by
\begin{eqnarray}
\varepsilon = \frac{M_p^2}{2}\left( \frac{U^{E}_\psi}{U^{E}}\right)^2\,, \quad\quad\eta = M_p^2\,\frac{U^{E}_{\psi\psi}}{U^{E}}\,,
\label{SR-parametersT}
\end{eqnarray}
where $U^{E}_{\psi}$ denotes derivative with respect to $\psi$, i.e., $U^{E}_\psi=dU^{E}/d\psi$ and $U^{E}_{\psi\psi}=d^{2}U^{E}/d\psi^{2}$. We obtain in terms of the field $\phi$
\begin{eqnarray}
\varepsilon &=&\frac{8 M_{p}^4}{\kappa ^2 \left(M_{p}^2+\xi  \phi ^2\right)^2}+\frac{16 \gamma M_{p}^2}{\kappa^2 \left(M_{p}^2+\xi  \phi ^2\right)}+O\left(\gamma ^2\right)\,,\\ \eta&=&-\frac{8 \left(M_{p}^2 \left(\xi  \phi ^2-2 M_{p}^2\right)\right)}{\kappa ^2 \xi ^2 \phi ^4}+\frac{32 \gamma M_{p}^2 \left(M_{p}^2+\xi  \phi ^2\right)}{\kappa ^2 \xi ^2 \phi ^4}+O\left(\gamma ^2\right)\,.\label{eetaT}
\end{eqnarray}
Inflation ends when $\varepsilon=1$ and we find
\begin{eqnarray}
\phi_{\rm end}&=& 2^{3/4} \sqrt{\frac{M_{p}^2}{\kappa  \xi }}+\frac{\left(2 \sqrt[4]{2}\right) \gamma  \sqrt{\frac{M_{p}^2}{\kappa \xi}}}{\kappa }+O\left(\gamma ^2\right)\\&=&\big(1.075+0.620 \gamma+O(\gamma)^2\big)\frac{M_p}{\sqrt{\xi}}\quad{\rm for}\quad\xi\gg 1\,.\label{mo3T}
\end{eqnarray}
The number of {\it e}-foldings during inflation is defined via
\begin{eqnarray}
N =\frac{1}{M^{2}_{p}}\int^{\psi_{\rm ini}}_{\psi_{\rm end}} \frac{U}{U_{\psi}}d\psi\simeq\int^{\phi_{\rm ini}}_{\phi_{\rm end}}\frac{\kappa ^2 \left(\frac{\xi\phi^2}{M_{p}^2}+1\right)}{4 \phi\left(\frac{\gamma \left(\xi\phi^2\right)}{M_{p}^2}+1\right)}d\phi\simeq \frac{\kappa^2}{8\gamma} \log \left(\frac{\gamma\xi\phi^2}{M_{p}^2}+1\right)\Bigg|^{\phi_{\rm ini}}_{\phi_{\rm end}}\,.
\end{eqnarray}
The above result can be combined with Eq.(\ref{mo3T}) allowing us to write:
\begin{eqnarray}
\phi_{\rm ini}&\simeq& \Bigg(2\sqrt{2}+\frac{4 \sqrt{2} \gamma  N}{\kappa ^2}+\frac{20 \sqrt{2} \gamma ^2 N^2}{3 \kappa ^4}\Bigg)\sqrt{\frac{N}{\kappa ^2}}\frac{M_{p}}{\sqrt{\xi }}\\&\simeq&\Big(8.94+178.89\gamma +2981.42\gamma^2\Big)\frac{M_{p}}{\sqrt{\xi }}\quad{\rm for}\quad N=60,\,\xi \gg 1\,.\label{ini3T}
\end{eqnarray}
It is evident that the $\gamma$-correction lifts up inflation to higher field values. Since we have assumed that inflation takes place in the large field regime, $\phi \gg {M_p /\sqrt{\xi}}$, it was found in Ref.\cite{Joergensen:2014rya} that $\xi\sim {\cal O}(10^{4})$ is required to generate the proper amplitude of density perturbations featuring a generic behavior of non-minimally coupled theories of single-field inflation. We can further determine field values to obtain $\phi_{\rm ini}\sim 0.1\,M_{p}$ for $\gamma \sim 0.01,\,\xi\sim 10^{4}$ and $N=60$. Interestingly, its value is sub-Planckian during inflation. However, with the large field assumption, we encounter the large field deviation since $\Delta \psi>M_{p}$ given by Eq.(\ref{psh}). Similar to $R^{2}$ and eternal inflation scenarios, it is in general a sense of difficulty of constructing models of large field inflation in string theory where the inflaton undergoes a super-Planckian excursion in field space. Nevertheless, a quantitative impression of this tension was suggested in Refs.\cite{Palti:2019pca,Matsui:2018bsy} where their analysis is applicable to that of our model. Therefore, we do not intentionally repeat it here and recommend the readers to follow those works. To generate the proper amplitude of the density perturbations the potential must satisfy at WMAP the normalization
condition \cite{Planck:2018jri}:
\begin{eqnarray}
\frac{U}{\varepsilon}\Big|_{\psi_{\rm ini}}=(0.0276 M_{p})^{4}
\end{eqnarray}
corresponding to the initial value assumed by the inflaton:
\begin{eqnarray}
\frac{4\lambda M_{p}^4 N^2}{3 \xi^2}-\frac{4 \gamma  M_{p}^4 \left(-12 \lambda  N^2 \log\left(\frac{2 M_{p} \sqrt{N}}{\sqrt{3}\Lambda\sqrt{\xi}}\right)+4 \lambda N^3+6 \lambda N^2\right)}{9 \xi ^2}\simeq (0.0276 M_{p})^4\,.
\end{eqnarray}
We therefore obtain
\begin{eqnarray}
\Lambda\simeq 0.7\sqrt{N} e^{\big(\frac{0.25}{\gamma }-\frac{1.1 \xi^2}{10^7 \left(\gamma  \lambda  N^2\right)}-0.33 N\big)}\frac{M_{p}}{{\sqrt{\xi}}}\,.
\end{eqnarray}
Ii is useful to write $\varepsilon$ and $\eta$ in terms of the number of e-foldings. Substituting Eq.(\ref{ini3T}) into Eq.(\ref{SR-parametersT}), we have
\begin{eqnarray}
\varepsilon \simeq \frac{3}{4 N^2}+\frac{\gamma }{N}\,,\quad
\eta \simeq -\frac{1}{N}+\frac{2\gamma}{3}\,.
\end{eqnarray}
The spectral index of curvature perturbation $n_s$ and the tensor-to-scalar
ratio $r$ are given in terms of the e-foldings $N$:
\begin{eqnarray}
n_{s} &=& 1-6\varepsilon+2\eta = 1-\frac{2}{N}-\frac{9}{2 N^2}+O\left(\gamma\right)\,,\\ r  &=& 16\varepsilon =\frac{12}{N^2}+\frac{16 \gamma }{N}+O\left(\gamma ^2\right)\,.
\end{eqnarray}
Here usual $\phi^{4}$ inflation refers to the results when setting $\gamma=0$, that is, non-minimally coupled $\phi^{4}$ inflation. Note that we have expanded results in $\gamma$ to clarify to what extend the results deviate from $\phi^{4}$ inflation. An expansion is, however, justified for tiny values of $\gamma$.

\section{Examination with the further refining swampland conjecture}\label{s3}
It is useful to define two new parameters for any scalar field $V(\phi)$
\begin{eqnarray}
F_{1}=\frac{|dV(\phi)/d\phi|}{V(\phi)}\,,
\label{f1}
\end{eqnarray}
and
\begin{eqnarray}
F_{2}=\frac{d^{2}V(\phi)/d\phi^{2}}{V(\phi)}\,.
\label{f2}
\end{eqnarray}
Considering Eq.(\ref{R12}), the above parameters can be recast in terms of the slow-roll parameters to yield
\begin{eqnarray}
F_{1}=\sqrt{2\varepsilon_{V}}\,,\quad F_{2}=\eta_{V}\,.
\label{f1f2}
\end{eqnarray}
Since $F_1$ and $F_2$ are written in terms of the slow-roll parameters, they can also be related to the spectrum index of the primordial curvature power spectrum $n_{s}$ and tensor-to-scalar ratio $r$. In the present case, it is rather straightforward to show that
\begin{eqnarray}
F_{1}=\sqrt{2\varepsilon_{V}}=\sqrt{\frac{r}{8}}\,,
\label{f11}
\end{eqnarray}
and
\begin{eqnarray}
F_{2}=\eta_{V}=\frac{1}{2}\big(n_{s}-1+3r/8\big)\,.
\label{f12}
\end{eqnarray}
Below we will consider three models of inflation and examine if they satisfy this new refined swampland conjecture, or not. In our analysis below, we constrain parameters of the models only for generic values of $\alpha>0$. 

\subsection{Inflation with E-form potential}
We can constrain values of $\alpha$ using a condition of $r<0.06$
\begin{eqnarray}
r\simeq\frac{12\alpha}{N^2}+\frac{18 \alpha ^2}{N^3}<0.06\,,
\end{eqnarray}
to obtain
\begin{eqnarray}
\alpha <0.0333 \sqrt{3N^3+100. N^2}-0.333N.
\end{eqnarray}
For $N=60$, we find $\alpha<13.5$. With the above result, we then choose $n_{s}=0.965$ and obtain from Eq.(\ref{ns1}): 
\begin{eqnarray}
\alpha=0.222222 N^2 \left(0.035\, -\frac{2}{N}\right).
\end{eqnarray}
Using $N=60$ yields $\alpha=1.333$, and therefore this model predicts $n_{s}\simeq 0.965$ and $r\simeq 0.004$ which are consistent with the observed data \cite{Planck:2018jri}. Inserting these values into Eq.(\ref{f1}) and Eq.(\ref{f2}), we obtain
\begin{eqnarray}
F_{1}&=&\sqrt{2\varepsilon_{V}}=\sqrt{\frac{r}{8}}=0.02236\,,\\F_{2}&=&\eta_{V}=\frac{1}{2}\big(n_{s}-1+3r/8\big)=-0.01675\,.
\label{f1f2}
\end{eqnarray}
Considering the refined swampland conjecture (\ref{R12}), we find
\begin{eqnarray}
c_{1}\leq 0.02236\quad{\rm or}\quad c_{2}\leq 0.01675\,.
\label{c1c21}
\end{eqnarray}
However, $c_{1}$ and $c_{2}$ are both not ${\cal O}(1)$, meaning that inflationary model with the E-form potential is in strong tension with the refined swampland conjecture. Let us examine whether the E-form model satisfies the refining swampland conjecture. Considering Eq.(\ref{ReRe}), we have
\begin{eqnarray}
(2\epsilon_{V})^{q/2}-a\eta_{V}\geq 1-a\,,\quad q>2\,.
\label{re111}
\end{eqnarray}
Substituting Eq.(\ref{c1c21}) into Eq.(\ref{re111}), we find
\begin{eqnarray}
0.02236^{q}+0.01675a\geq 1-a\,\quad{\rm or}\quad 0.02236^{q}\geq 1-1.01675\,a\,.
\label{re111}
\end{eqnarray}
If we can find $a$ to satisfy the condition
\begin{eqnarray}
\frac{1}{1.01675}(1-0.02236^{q})\leq a<1\,,
\quad q>2,
\label{r111}
\end{eqnarray}
then the further refining swampland conjecture can be satisfied. In this case, when $a=\nicefrac{1}{1.01675}$, we have $1-1.01675\,a=0$. Therefore, we can examine that when $a < \nicefrac{1}{1.01675}$, we can always find a $q$ whose value is larger than $2$. It is possible to give an example of values of the parameters $a,\,b,\,q$, which work for this model. From Eq.(\ref{r111}), we use $q=2.2$ which is satisfied by a condition $q>2$. We find for this particular case that $0.983296 \leq a<1$ and choose $a=0.9834<\nicefrac{1}{1.01675}=0.983526$ and $1-a=1-0.9834=b=0.0166>0$.

\subsection{Inflation with T-form potential}
We can constrain values of $\alpha$ using a condition of $r<0.06$
\begin{eqnarray}
r\simeq\frac{12\alpha}{N^2}<0.06\,,
\end{eqnarray}
to obtain
\begin{eqnarray}
\alpha <0.005 N^2.
\end{eqnarray}
For $N=60$, we find $\alpha<18$. With the above result, we then choose $n_{s}=0.965$ and obtain from Eq.(\ref{ns1}): 
\begin{eqnarray}
\alpha=0.666667 N^2 \left(0.035\, -\frac{2}{N}\right).
\end{eqnarray}
Using $N=60$ yields $\alpha=4$, and therefore this model predicts $n_{s}\simeq 0.965$ and $r\simeq 0.0133$ which are consistent with the observed data \cite{Planck:2018jri}. Inserting these values into Eq.(\ref{f1}) and Eq.(\ref{f2}), we obtain
\begin{eqnarray}
F_{1}&=&\sqrt{2\varepsilon_{V}}=\sqrt{\frac{r}{8}}=0.040825\,,\\F_{2}&=&\eta_{V}=\frac{1}{2}\big(n_{s}-1+3r/8\big)=-0.0150\,.
\label{f1f2}
\end{eqnarray}
Considering the refined swanpland conjecture (\ref{R12}), we find
\begin{eqnarray}
c_{1}\leq 0.040825\quad{\rm or}\quad c_{2}\leq 0.0150\,.
\label{c1c22}
\end{eqnarray}
However, neither $c_{1}$ nor $c_{2}$ is positive constant of the order of ${\cal O}(1)$, meaning that inflationary model with the E-form potential is in strong tension with the refined swampland conjecture. Let us examine whether the E-form model satisfies the refining swampland conjecture. Considering Eq.(\ref{ReRe}), we have
\begin{eqnarray}
(2\epsilon_{V})^{q/2}-a\,\eta_{V}\geq 1-a\,,\quad q>2\,.
\label{re22}
\end{eqnarray}
Substituting Eq.(\ref{c1c22}) into Eq.(\ref{re22}), we find
\begin{eqnarray}
0.040825^{q}+0.0150\,a\geq 1-a\,.
\label{re111}
\end{eqnarray}
If we can find $a$ to satisfy the condition
\begin{eqnarray}
\frac{1}{1.0150}(1-0.040825^{q})\leq a<1\,,
\quad q>2,
\label{re112}
\end{eqnarray}
then the further refining swampland conjecture can be satisfied. For the second model, when $a=\nicefrac{1}{1.0150}$, we have $1-1.0150\,a=0$. Hence, we can examine that when $a < \nicefrac{1}{1.0150}$, we can always find a $q$ whose value is larger than $2$. Similarly, we can give an example of values of the parameters $a,\,b,\,q$, which work for this model. From Eq.(\ref{re112}), we use $q=2.2$ which is satisfied by a condition $q>2$. We find for this particular case that $0.984356 \leq a<1$ and choose $a=0.9844<\nicefrac{1}{1.0150}=0.985222$ and $1-a=1-0.9844=b=0.0156>0$.
\subsection{Quantum corrected inflation}
We can constrain values of $\alpha$ using a condition of $r<0.06$
\begin{eqnarray}
r\simeq\frac{12}{N^2}+\frac{16 \gamma }{N}<0.06\,,
\end{eqnarray}
to obtain
\begin{eqnarray}
\gamma <\frac{0.00375}{N}\left(N^2-200\right).
\end{eqnarray}
For $N=60$, we find $\gamma<0.2125$. Using $N=60,\,\gamma=0.10$, this model predicts $n_{s}\simeq 0.965$ and $r\simeq 0.030$ which are consistent with the observed data \cite{Planck:2018jri}. Inserting these values into Eq.(\ref{f1}) and Eq.(\ref{f2}), we obtain
\begin{eqnarray}
F_{1}&=&\sqrt{2\varepsilon_{V}}=\sqrt{\frac{r}{8}}=0.061237\,,\\F_{2}&=&\eta_{V}=\frac{1}{2}\big(n_{s}-1+3r/8\big)=-0.011875\,.
\label{f1f2}
\end{eqnarray}
Considering the refined swanpland conjecture (\ref{ReRe}), we find
\begin{eqnarray}
c_{1}\leq 0.061237\quad{\rm or}\quad c_{2}\leq 0.011875\,.
\label{c1c23}
\end{eqnarray}
Neither $c_{1}$ nor $c_{2}$ are of the order of ${\cal O}(1)$ which mean that inflationary model with the quantum corrected potential is in strong tension with the refined swampland conjecture. Let us examine if this model satisfies the refining swampland conjecture. Considering Eq.(\ref{ReRe}), we have
\begin{eqnarray}
(2\epsilon_{V})^{q/2}-a\eta_{V}\geq 1-a\,,\quad q>2\,.
\label{re333}
\end{eqnarray}
Substituting Eq.(\ref{c1c23}) into Eq.(\ref{re333}), we find
\begin{eqnarray}
0.061237^{q}+0.011875\,a\geq 1-a\,.
\label{re1}
\end{eqnarray}
If we can find $a$ to satisfy the condition
\begin{eqnarray}
\frac{1}{1.011875}(1-0.061237^{q})\leq a<1\,,
\quad q>2,
\label{re113}
\end{eqnarray}
then the further refining swampland conjecture can be satisfied. When $a=\nicefrac{1}{1.011875}$, we see that $1-1.011875\,a=0$. Hence, we can examine that when $a < \nicefrac{1}{1.011875}$, we can always find a $q$ whose value is larger than $2$. In the last model, we can also give an example of values of the parameters $a,\,b,\,q$, which work for our model. From Eq.(\ref{re113}), we use $q=2.2$ which is satisfied by a condition $q>2$. We find for this particular case that $0.986145 \leq a<1$ and choose $a=0.9862 <\nicefrac{1}{1.011875}=0.988264$ and $1-a=1-0.9862=b=0.0138>0$. Interestingly, for this model the scale $\Lambda$ can be constrained. For instance, substituting $N=60,\,\lambda=\nicefrac{1}{4},\,\gamma=0.01504$ and $\xi=10^{4}$, we find that 
\begin{eqnarray}
\Lambda\simeq 0.7\sqrt{N} e^{\big(\frac{0.25}{\gamma }-\frac{1.1 \xi ^2}{10^7 \left(\gamma  \lambda  N^2\right)}-0.33 N\big)}\frac{M_{p}}{{\sqrt{\xi}}} \sim 10^{-3}\,M_{p}\simeq [0.245-1.22]\times 10^{16}\,{\rm GeV}\,,
\end{eqnarray}
with $\lambda=\nicefrac{1}{4}$ being a standard value \cite{Joergensen:2014rya}. This value is close to the typical grand unification scale $M_{\rm GUT}$ of $10^{16}$\,GeV with the lower value obtained for the reduced Planck mass of $2.44\times10^{18}$\,GeV and the higher one for the standard one of $1.22\times10^{19}$\,GeV.

\section{Conclusion}\label{conclud}
In the present work, we have found that the inflation models in a general scalar-tensor theory with exponential and hyperbolic tangent forms of potential as well as model with quantum corrected potential are all in strong tension with the refined de Sitter swampland conjecture. In other words, we have demonstrated in all three models that positive values $c_{1}$ and $c_{2}$ are both not ${\cal O}(1)$. Similarly, notice that Palatini inflation model is also in strong tension with the refined de Sitter swampland conjecture \cite{Liu:2021qsr}. 

However, we have further tested if these three models of inflation can satisfy this further refining de Sitter swampland conjecture or not. Regarding our analysis, we have discovered that these three inflationary models can always satisfy this new refined swampland conjecture if only we adjust the relevant parameters $a,\,b=1- a$ and $q$. Therefore, the three inflationary models might all be in “landscape” since the “further refining de Sitter swampland conjecture” is satisfied.

Nevertheless, the upper and lower bounds of these three parameters $a,\,b=1-a$ and $q$ using this new swampland conjecture can not be quantified. In the future work, to constrain the range of these physical parameters, new other swampland conjecture in string theory may be worth investigating.

\section*{Acknowledgements}
P. Channuie acknowledged the Mid-Career Research Grant 2020 from National Research Council of Thailand (No.NRCT5-RSA63019-03) and is partially supported by the National Science, Research and Innovation Fund (SRF) with grant No. P2565B202.

\subsection*{CRediT authorship contribution statement}
{\bf Jureeporn Yuennan}: Writing – review \& editing. {\bf Phongpichit Channuie}: Formal analysis, Methodology, Writing – original draft, Writing – review \& editing.

\subsection*{Conflict of Interest}
The authors declare no conflict of interest.

\end{document}